\documentclass[12pt]{article}
\usepackage{epsfig, bm, amsfonts, amssymb}
\usepackage{multirow}
\usepackage{tabularx}
\usepackage{hhline}
\usepackage[normalem]{ulem}
\usepackage{cite}
\usepackage[none]{hyphenat}
\usepackage[textsize=tiny]{todonotes}
\usepackage{chngcntr}
\usepackage[greek,english]{babel}
\usepackage{teubner}
\usepackage[normalem]{ulem}
\usepackage{amsmath}
\usepackage{soul}

\counterwithin{figure}{section}
\counterwithin{table}{section}

\RequirePackage{color}

 \definecolor{MyDarkGreen}{rgb}{0.02,0.60,0.06}

\date{15 May 2017}

\def\q{{\hbox{\foreignlanguage{greek}{\coppa}}}}
\def\qq{{\hbox{\foreignlanguage{greek}{\footnotesize\coppa}}}}

\title{{\bf Universal Finite-Size Scaling for Percolation Theory in High Dimensions}}
\author{ 
 {\it Ralph Kenna}$^{1,*}$ and
 {\it Bertrand Berche}$^{2,*}$ \\~\\
\begin{tabular}{ll}
  $^1$     & {\small{Applied Mathematics Research Centre, Coventry University, CV1 5FB, England}}\\
	$^2$		 & {\small{Universit\'e de Lorraine, Statistical Physics Group, IJL, UMR CNRS 7198, }}\\
	         & {\small{Campus de Nancy, B.P. 70239, 54506 Vand\oe uvre l\`es Nancy Cedex, France}}\\
  $^*$     & {\small{${\mathbb L}^4$ Collaboration \& Doctoral College for the Statistical Physics of Complex Systems,}} \\       
	         & {\small{Leipzig-Lorraine-Lviv-Coventry, Europe}} 
\end{tabular}
{}\\~\\
}

\begin{document}
\maketitle
\vspace{-1cm}
{\Large
  \begin{abstract}
%
\noindent
We present a unifying, consistent, finite-size-scaling picture for percolation theory bringing it into the framework of a general, renormalization-group-based, scaling scheme for systems above their upper critical dimensions $d_c$.
Behaviour  at the critical point is non-universal in $d>d_c=6$ dimensions. 
Proliferation of the largest clusters, with fractal dimension $4$, is associated with the breakdown of hyperscaling there when free boundary conditions are used.
But when the boundary conditions are periodic, the maximal clusters have dimension $D=2d/3$, and obey random-graph asymptotics.
Universality is instead manifest at the pseudocritical point, where the   failure of hyperscaling in its traditional form  is universally associated with   random-graph-type asymptotics for critical cluster sizes, independent of boundary conditions.

                        \end{abstract} }
%
  \thispagestyle{empty}
%
%
  \newpage
%
                  \pagenumbering{arabic}

\section{Introduction}
\label{Introduction}
\setcounter{equation}{0}


Percolation theory has been the subject of extensive mathematical and simulational studies and is of relevance in a broad range of fields including physics,  chemistry, network science, sociology, epidemiology, and geology \cite{Essam,Stauffer,CM,ArGr14,Sa15}.
The existing statistical-mechanical picture of scaling close to the critical point  is as follows.
Below the upper critical dimension $d_c$, where hyperscaling holds in its standard form, the average number of incipient spanning clusters is independent of finite-lattice size \cite{Essam,Stauffer}.
The upper critical dimensionality is $d_c =6$, above which mean-field theory holds as a  predictor of  exponents governing critical behaviour \cite{MFTperc,AhGe84,AiNe84,HaSa90,BaAi91,HaSl94,ShRo02,HaHo03,Ha08,FiHo15,HeHo16}.
Above $d_c$, although there can only be a unique infinite cluster in the thermodynamic limit \cite{AiKeNe87,BuKe89}, the 
{spanning clusters} on finite lattices may proliferate and this phenomenon is associated with the breakdown of hyperscaling in its traditional form~\cite{Conig85,Aizen97,Co00}.
This picture was  formed over 30 years ago \cite{Conig85} in parallel with the formulation of the scaling picture for thermal phase transitions~\cite{FiHa83,BNPY}.
More recently it has been proven that scaling in high dimensions is dependent on boundary conditions at the infinite-volume  critical point.
While  proliferation  of the largest clusters is expected for free boundary conditions (FBCs), 
where  they have fractal dimension $D=4$  \cite{Aizen97},  
rigorous proofs have established that, under certain conditions, $D=2d/3$ for other  boundary conditions, such as periodic boundary conditions (PBCs), $d$ being the dimension of the underlying lattice
\cite{BoCh05a,BoCh05b,HeHo07,HeHo11,Pete}. 
This is known as {\emph{random-graph asymptotics}} because, with $V=L^d$, it is similar to the behavior of the largest critical cluster on the complete graph of $V$ sites \cite{HeHo07,HeHo16}.

Thus the range of validity of the original finite-size scaling (FSS) picture of percolation in high dimensions has been  revised in favour of another, which is not universal at the critical point.
The original picture was built on renormalization-group concepts such as  Kadanoff's length rescaling~\cite{Ka66}, Fisher's dangerous irrelevant variables~\cite{FiHa83}, as well as Binder's thermodynamic length~\cite{Bi85}.
The question therefore arises as to the status of these concepts under the new picture.
The non-universality of FSS at the critical point also brings into question the  mechanism for the failure of hyperscaling under different boundary conditions.
Additionally, there are issues associated with {incipient spanning} and {infinite} clusters \cite{Aizen97,Aizen96} and there are apparent discrepancies between theory and simulations \cite{FoStCo04}.
These need to be explained in a manner compatible with the scaling picture, the foundation for which is the renormalization group.
That picture for percolation in high dimensions was reviewed in Refs.\cite{St97,Aizen97,Co00,FoStCo04}
and summarised in Ref.\cite{FoAhCoSt04} as ``incomplete, leaving an open challenge for future research.''
Here we seek to meet that challenge.

Recently, we introduced a new formulation for scaling  and finite-size scaling above the upper critical dimension~\cite{ourNPB,ourCMP,ourEPL,ourEPJB,ourreview,ourPRL}.
The new framework is variously termed $Q$-scaling or $Q$-FSS or simply $Q$-theory to distinguish it from the older scaling scenario and its  modifications~\cite{Brankov}. 
In Refs.\cite{ourNPB,ourCMP,ourEPL,ourEPJB,ourreview,ourPRL}, the new set-up was expressed in terms of $\phi^4$-theory and tested for spin models.
Here we show that extending $Q$-scaling to percolation theory above the upper critical dimension  resolves a number of important puzzles and unifies the above pictures. 

In particular, it shows that the maximal-cluster-proliferation picture only holds at the infinite-volume critical point for systems with FBCs where it is associated with the breakdown of hyperscaling in its standard form.
For PBCs, the number of maximal  clusters is independent of lattice volume and the failure of hyperscaling there is due to the sizes of critical clusters scaling as a larger power of the  lattice extent.
This picture also holds at the pseudocritical point irrespective of boundary conditions and this is where universality is located.
(The pseudocritical point is defined, e.g., by the  activity probability for which the correlation length of a finite-size system takes its maximum value.)
The scaling predictions  at the pseudocritical point in the FBC case  are a new consequence of the $Q$-picture.

In the next section, we briefly review the evolution of percolation theory above six dimensions.
We discuss how it parallels the progress of $\phi^4$-theory above $d=4$ and we summarize the recent development of $Q$-scaling there \cite{ourNPB,ourCMP,ourEPL,ourEPJB,ourreview,ourPRL}.
Then, in Section~3,  we integrate percolation scaling  theory~\cite{Conig85,Co00,Aizen97} into  the $Q$-framework, and show that it also accommodates the more recent results of Refs.\cite{IIF2,BoCh05b,BoCh05a,HeHo07,HeHo11,HoSa14,HaHo03,Ha08,Pete,FiHo15,NaPe08,IIF3,HeHo16}. 
In Section~\ref{Conclusions} we discuss the status of concepts and resolution of puzzles mentioned above and we conclude by making  proposals for future research.

\section{Scaling theories for percolation in high dimensions}
\label{History}

We consider  percolation  on lattices of linear extent $L$ and volume $V=L^d$.
The probability that a site or bond is occupied is $p$.
For small values of $p$, one typically has many small clusters of occupied sites or bonds and no infinite cluster in the thermodynamic limit.
For sufficiently large $p$, on the other hand, there is typically also a {spanning cluster} which becomes infinite in size as $L$ becomes infinite. 
The  pseudocritical point is  characterised by $p_L$ in finite volume and the critical point in the infinite-volume limit is $p_\infty$.
There, it is  believed that there is no infinite cluster precisely at $p_\infty$~\cite{HaSa90,HaSl94}.
The extension to long-range or spread-out models is straightforward \cite{ourEPJB,ourPRL}.

At the critical point, in the infinite-volume limit, the density of clusters of size or mass $s$ (the number of sites occupied by the cluster) follows a power law, scaling as $s^{-\tau}$.
The  size of the largest cluster diverges there as  $|p-p_\infty|^{-1/\sigma}$ 
and the mean radius of the remaining clusters (the correlation length) diverges as $|p-p_\infty|^{-1/\nu}$.
In the counterpart FSS  window, whose width is $L^{-\lambda}$, the correlation length 
scales as  $L^{\qq}$ and the $L^X$ largest clusters scale in size  as $L^D$.

\subsection{Review of percolation theory in high dimensions}
\label{History1}

Prior to the 1980s, it was generally believed that there can only be a single infinite cluster at the percolation threshold \cite{Essam,deArc87}.
Then Newman and Schulman showed rigorously that the number of infinite clusters in infinite volume could actually be zero, one or infinity \cite{NeSc81,NeSc81b}. 
In particular, in high dimensions, an infinite number of sparse (zero-density) infinite clusters could, in principle, exist.
Such clusters could be ``rough'' or ``filamentary'' and they avoid meeting each other in high dimensions~\cite{NeSc81,NeSc81b}.
It was expected, however, that at most a single infinite cluster exists in dimensions below $d_c$.

In 1984, Aharony, Gefen and Kapitulnik gave an argument linking the behaviour of percolation clusters to dangerous irrelevant variables and the breakdown of hyperscaling above the upper critical dimension~\cite{AhGe84}.
They claimed that the fractal dimensionality  of the {infinite cluster} at the percolation threshold for dimensionalities $d>6$ is stuck at $D=4$ rather than the naive finite-size scaling prediction $D=d-2$ which comes from inserting mean-field critical exponents into the $d<d_c$ scaling predictions. 

In 1985, Coniglio  incorporated these developments to give a geometrical interpretation for the breakdown in hyperscaling above $d_c$~\cite{Conig85}.
According to this theory, the validity of hyperscaling below $d_c$ is equivalent to  the presence of only a single critical cluster. 
Its breakdown in high dimensions is then supposed to be related to the proliferation and interpenetration of large clusters, and to the appearance at criticality of an infinite number of infinite clusters of zero density. 
For the system to accommodate a diverging number of distinct critical clusters they have to bypass each other and can interweave as they do so; proliferation and interpenetration come hand in hand in Coniglio's picture~\cite{Conig85,Co00}.
In finite volume, the theory predicts there are  $L^{d-6}$ such spanning clusters in the critical region, each with a mass $L^D$ with fractal dimension $D = 4$.

An early numerical study of the topic came in 1987 when de Arcangelis  provided evidence that  there is indeed a unique critical cluster spanning the whole system at $p_\infty$ for $ d< d_c$, but  multiple fragile spanning clusters when $d > d_c$~\cite{deArc87}. 
The numerics for small sizes did not quite match up with quantitative predictions, however. 
Ray and Klein numerically verified some predictions in the thermodynamic limit for long-range bond percolation  \cite{RaKl88}.
In 1995 Aharony and Stauffer extended the scaling theory to consider the shift in the percolation threshold from the critical point $p=p_\infty$ in infinite volume to the pseudocritical point $p_L$ in finite volume~\cite{AhSt95}. 
Their theoretical considerations were based on the arguments of Ref.\cite{BNPY} with  new length scales introduced into the problem by dangerous irrelevant variables \cite{FiHa83}. 
In the terminology of spin models, the relevant new length scale is the thermodynamic length \cite{AhGe84,Bi85}.
They predicted $p_\infty - p_L \sim AL^{-2} + BL^{-d/3}$ with $A$ vanishing for PBCs~\cite{AhSt95}.
This prediction received some numerical support in Ref.\cite{StZi00} where a positive value for $A$ was measured for mixed boundary conditions. 
(The precise value of $A$ is not required to determine the pseudocritical point, however, since, as mentioned above, it can be obtained by maximising  the correlation length or, indeed, the mean cluster size.) 
Since the rounding is given by $L^{-d/3}$ for both PBCs and FBCs, but the shifting is $L^{-2}$ in the latter case, the rounding is smaller than the shifting there.

The apparent demarcation between $d<d_c$ and $d>d_c$, in that the former manifests a single {spanning} cluster while the latter can have an infinite number of them, was undermined in 1996 when Hu, Lin and 
Hsu produced numerical evidence for {numerous spanning clusters} in two dimensions~\cite{Hu96,Hsu01}.
Aizenman then gave a non-zero  probability for $k > 1$ {spanning clusters} in two dimensions~\cite{Aizen97}. 
That there can be more than one spanning cluster at $p_\infty$ for $d<d_c$ has since been confirmed by Sen, Shchur, Cardy and others in the late 1990s~\cite{Se96,Se97,Lev97,Lev98,St97,Cardy98,Jan98,Sh00}.

However, in Refs.\cite{AiKeNe87,BuKe89} it was shown that  when there is percolation ($p>p_\infty$), the  infinite cluster (with non-zero density) has to be unique.
The single dominant cluster at $p=p_\infty$ is sometimes referred to as the {\emph{{incipient} infinite cluster}}~\cite{IIF1,IIF2,IIF3}.
Aizenman also produced a number of rigorous results confirming the picture of Refs.\cite{AhGe84,Conig85} for $d > 6$ \cite{Aizen97}. 
He considered percolation in a box of width $L$ under bulk boundary conditions, meaning that the system is viewed as part of a larger domain; clusters in the box can be connected through sites outside of it.
Assuming vanishing anomalous dimension above the upper critical dimension (established in certain circumstances in Refs.\cite{HaSa90,HaHo03,Ha08,FiHo15}), 
it was found that there are typically $L^{d-6}$ {\emph{{spanning clusters}}} of size $L^4$.
This was in line with  the renormalization theory \cite{AhGe84,Conig85} and  similar scaling was expected for FBCs \cite{Aizen97}.
A different scaling behaviour was conjectured for the PBC case~\cite{Aizen97}, namely $D=2d/3$, which corresponds to  the  scaling law for the complete graph \cite{ErRe60}.

Indeed, it had long been known that at a certain probability, the size of the largest component of the  Erd{\H{o}}s-R{\'{e}}nyi random graph of $V$ sites is $V^{2/3}$.
This was proven by Bollob{\'{a}}s \cite{Bo84} and {\L}uczak \cite{Lu90} who also found that the scaling window has size $V^{-1/3}$. 
See also Ref.\cite{Aldous} where the scaling of the largest clusters for percolation on the complete graph is proved in detail.
If percolation on finite graphs is similar to that in a Erd{\H{o}}s-R{\'{e}}nyi random graph, it would mean that the largest cluster has fractal dimension  $D={2d/3}$ and that this holds inside a scaling window given by  $\lambda = d/3$.
This was the scaling window derived by Aharony and Stauffer in Ref.\cite{AhSt95} (see also Ref.\cite{St94}) in the case of PBCs.

In Refs.\cite{BoCh05a,BoCh05b}, the connection with the Erd{\H{o}}s-R{\'{e}}nyi random graph was further investigated for percolation on  complete graphs, tori,  hypercubes as well as finite-range percolation.
Assuming that a quantity called the triangle diagram is sufficiently small (to ensure that the model corresponds to the mean-field theory)  the largest connected component was found to indeed scale as $V^{2/3}$ inside a window  of width $V^{-1/3}$ about the pseudocritical point where the expected cluster size is $V^{1/3}$.
This was further developed for the torus in Refs.\cite{HeHo07,HeHo11}, where it was shown that the critical point of the infinite lattice lies inside this scaling window.
According to this picture, the diameter of large clusters  is of order $V^{1/3}$ \cite{NaPe08,HoSa14,KoNa11}.
The similarity between percolation on high-dimensional systems with PBCs and that of  Erd{\H{o}}s-R{\'{e}}nyi, gave rise to the term ``random-graph asymptotics'' \cite{HoSa14,HeHo16}.\footnote{Here and throughout we use the term ``random-graph asymptotics'' for systems (i) whose critical cluster sizes scale in the same way as on the Erd{\H{o}}s-R{\'{e}}nyi random graph and (ii) where the window for such scaling has width $\sim V^{1/3}$. If (i) is satisfied but not (ii), we employ the looser term ``random-graph-type asymptotics''.}

In Ref.\cite{HoSa14}, van der Hofstad and Sapozhnikov investigated the cycle structures of the critical clusters for percolation on tori in high dimensions.
(Any cycle having displacement at least a fraction of $L$, for example, may be considered long \cite{HeHo16}.)
They showed that long cycles, of mass at least $\sim L^{d/3}$,  may be non-contractible in that, when considered as continuous curves, they can't be deformed to a point. 
In other words they wind around the torus at least once.
Actually, van der Hofstad and Sapozhnikov showed that such long cycles  can go around the torus $L^{d/d_c-1}$ times \cite{HoSa14,HeHo16}.
This claim was repeated on a numerical basis in a related context in Ref.\cite{Grimm}.

To summarise, on the one hand we have the  Kadanoff-scaling (based ultimately on the renormalization group) and geometrical theories which tell us that $D=4$ and $X=d-6$ above the upper critical dimension \cite{Conig85,Co00}.
These results were proven for bulk boundary conditions at criticality when the correlation function decays with vanishing anomalous dimension  \cite{Aizen97}, and this in turn is a prediction of mean-field theory \cite{HaSa90,HaHo03,Ha08,FiHo15}. 
In this (Kadanoff-scaling) context, Aharony and Stauffer give that $\lambda = d/3$ and $\lambda = 2$ for PBCs and FBCs, respectively~\cite{AhSt95}.
On the other hand  the suggestion from the Erd{\H{o}}s-R{\'{e}}nyi picture is that $D=2d/3$ and that $\lambda = d/3$.
This has been proven rigorously, subject to certain conditions, at least for PBCs in Refs.\cite{BoCh05a,BoCh05b,HeHo07,HeHo11}.

A number of questions arise.
Since the Kadanoff renormalization picture, which associates the breakdown of hyperscaling with Coniglio's proliferation scenario, manifestly only applies to the situation of FBCs at $p_\infty$, what is the counterpart renormalization-group explanation for hyperscaling collapse for PBCs?
What FSS applies at $p_L$ in the FBC case?
What are the values of the other critical exponents $\tau$, $\sigma$ and $X$?
Hyperscaling notwithstanding, what is the reason for the apparent breakdown of universal FSS in percolation theory above the upper critical dimension?
Numerical simulations have hitherto delivered unsatisfactory results.
What is a sensible direction for future simulations aimed at testing the various scenarios?

In what follows, we provide answers to all of these questions and more. 
Our approach follows the recent developments of Refs.\cite{ourNPB,ourCMP,ourEPL,ourEPJB,ourreview,ourPRL}.
To explain and contextualise these, we have to give a brief review of $\phi^4$-theory, which has evolved in parallel to percolation theory, and of recent developments there.

\subsection{Universal and non-universal FSS for $\phi^4$-theory in high dimensions}
\label{History2}

To place the theoretical problems associated with percolation in high dimensions in the context of this paper, we look at the parallel situation in spin models above~$d_c$. 
For the scalar $\phi^4$-theory, $d_c=4$ and the theory may be formulated in terms of an expansion of the free energy density in powers of the local order parameter $\phi({\bf x})$,
\begin{equation}
F[\phi]=\int d^dx \left({{\textstyle\frac 12} t \phi^2({\bf x})+ {\textstyle\frac 14}u\phi^4({\bf x}) -h\phi({\bf x})+{\textstyle\frac 12}|{\boldmath\nabla}\phi|^2 }\right),
\label{eqLandauExpansion}\end{equation} 
where $t$ is proportional to the deviation from the critical temperature and $h$ from the critical magnetic field.

Dangerous irrelevant variables, which were integral to the percolation theory developed by Aharony et al. \cite{AhGe84}, were first introduced by Fisher for infinite-volume spin systems in Ref.\cite{FiHa83}.
The theory was formulated to repair a mismatch between some of the critical exponents coming from the Gaussian fixed point and those predicted by  mean-field theory or Landau theory. These were the exponents $\alpha$ for the specific heat; $\beta$  for the temperature dependence of the magnetisation and $\delta$ for its dependency on external field. In  mean-field $\phi^4$-theory, they take the values $\alpha=0$, $\beta=1/2$ and $\delta=3$, while they have dimension-dependent values at the Gaussian fixed point, $\alpha^{\rm{G}} = 2-d/2$, $\beta^{\rm{G}} = (d-2)/4 $ and $\delta^{\rm{G}} =(d+2)/(d-2) $.
Each of these is derivable from the free energy (\ref{eqLandauExpansion}), and the danger is associated with the presence of the irrelevant scaling field $u$ there.
Fisher's scenario therefore applies to this sector.
Because the values of the correlation-length exponent $\nu^{\rm{G}}=1/2$ and the anomalous dimension $\eta^{\rm{G}}=0$ coming from the Gaussian fixed point coincide with those from Landau theory, it was not expected that the irrelevant term presents danger for  the correlation sector. 
Since the susceptibility is linked to the correlation function by the fluctuation-dissipation theorem, one expects it to escape the danger of irrelevant variables too.
Indeed, the value $\gamma^{\rm{G}} = 1$ coming from the Gaussian fixed point is the same as that from mean-field theory, and this seemed to confirm said belief.

In 1985, Binder, Nauenberg, Privman and Young applied Fisher's concept of dangerous
irrelevant variables to the finite-volume case of the $\phi^4$-theory~\cite{BNPY}.
Crucially, their theory demanded that the correlation length,  which we express as $\xi(L^{-1})$, cannot exceed the actual length $L$ of the system, an assumption explicitly inputted on physical grounds.
(Here, and throughout, we express size dependency in terms of $L^{-1}$ instead of $L$ so that the corresponding argument vanishes, rather than diverges, in the thermodynamic limit.)
A consequence was that the standard formulation of FSS in terms of a ratio of length scales $\xi (0)/L$ failed to deliver the correct size dependency of thermodynamic functions above the upper critical dimension.
To achieve this, Binder introduced another length scale: the thermodynamic length~\cite{Bi85}. 
Secondly, not only did hyperscaling and FSS break down above $d_c$ but universality did too; the finite-size behaviour seemed to depend strongly on boundary conditions.
In particular, free boundaries appeared to deliver different results than periodic boundaries~\cite{ourNPB,RuGa85,LuMa11,LuMa14}. 

The standard paradigm was developed in a rather ad-hoc manner.
Although it delivered the correct exponents by construction, it had a number of shortcomings.
For example, it was unexplained a priori why the dangerous-irrelevant-variable mechanism would apply to one sector and not the other and, the Ginzburg criterion notwithstanding, there was no explanation for the collapse of universality above $d_c$.
The theory also involved a distinction between the behaviour of correlation functions over ``short long'' distances and ``long long'' distances which has parallels in percolation theory.

Under the new $Q$-scheme \cite{ourNPB,ourCMP,ourEPL,ourEPJB,ourreview,ourPRL}, the thermodynamic length is obviated and hyperscaling (in a new form), finite-size scaling as well as  universality all  hold above the upper critical dimension. 
A crucial ingredient that differentiates $Q$-theory from the earlier set-up is that dangerous irrelevant variables are also allowed to affect the correlation sector of the model.
In particular, the restriction that the correlation length be bounded by the length no longer applies~\cite{Br82}.
This necessitates the introduction of a new exponent governing the FSS of the correlation length:
\begin{equation}
 \xi(L^{-1}) \sim L^{\qq} .
 \label{koppa}
\end{equation}
The new exponent $\q$ (``koppa'') is properly called a pseudocritical exponent rather than a critical one since it controls FSS inside  a scaling window rather than scaling in infinite volume with temperature or magnetic field.
It takes the value $\q = 1$ for $d<d_c$ where the standard theory of FSS holds sway. 
However, above the upper critical dimension, $\q = d/d_c$ (at least at pseudocriticality) so that the correlation length scales faster than the length of the system.
For spin models with PBCs, the FSS window, in which Eq.(\ref{koppa}) holds, includes both the critical and pseudocritical points. 
For FBCs, on the other hand, it includes only the pseudocritical point;
because of the largeness of the shifting the critical point lies outside the region in which Eq.(\ref{koppa}) holds~{\cite{RuGa85,ourNPB,ourEPL,ourCMP,ourreview,ourEPJB,ourPRL,WiYo14,Italians}.}
Thus $\q$ is physical and universal, like the other  exponents.
The new form for hyperscaling, which holds in all dimensions is \cite{ourNPB}
\begin{equation}
 \frac{\nu  d}{\q}  =2-\alpha .
 \label{hyperscalingclusters}
 \end{equation}
The breakdown in the old version of hyperscaling ($\nu d = 2 - \alpha$), at least for thermal phase transitions,  is because $\q \ne 1$ for $d>d_c$.
These results are consistent with explicit calculations and numerics in the case of the $\phi^4$-theory~{\cite{RuGa85,ourNPB,ourEPL,ourCMP,ourreview,ourEPJB,ourPRL,WiYo14,Italians}.}
Moreover, the old picture, in which $\xi (L^{-1})$ is bounded by $L$, is by now firmly falsified \cite{ourreview}.

The discussion above, and the difference between scaling with PBCs and FBCs, can 
be rendered explicit using the so-called homogeneity assumption for the free energy density and the correlation length for a system of size $L$ at reduced temperature $t$ and in reduced field $h$:  
\begin{eqnarray}
f(L^{-1},t,h) &=& b^{-d} F (bL^{-1},b^{y_t}t,b^{y_h}h,b^{y_u}u), 
\label{B11}\\
 \xi(L^{-1},t,h) &=& b    \Xi (bL^{-1},b^{y_t}t,b^{y_h}h,b^{y_u}u).
\label{B22}
\end{eqnarray}
The effect of the dangerous irrelevant variable above $d_c$ is to change  the renormalization-group eigenvalues of the reduced thermal and magnetic scaling fields, $t$ and $h$,  
from their Gaussian values
\begin{equation}
y_t^{\rm G}=2\quad{\rm and}\quad y_h^{\rm G} (d)=\frac d2+1,
\label{Eq-yGaussian}
\end{equation}
to $y_t^*$ and $y_h^*$, respectively, so that~\cite{ourNPB,BrZJ85}
\begin{eqnarray}
    {f(L^{-1},t,h)} &=& b^{-d} F(bL^{-1},b^{y_t^*}t,b^{y_h^*}h)
   \label{eqfHomogeneity} 
   \\
   {\xi (L^{-1},t,h)} &=& b^{\qq} \Xi (bL^{-1},b^{y_t^*}t,b^{y_h^*}h)
	 \label{eqXiHomogeneity}
\end{eqnarray}
where  
\begin{equation}
y_t^* (d)=\q y_t^{\rm G} =\frac{\q}{\nu}=2\q=\frac{d}{2},
\quad{\rm and}
\quad 
y_h^* (d)=\q y_h^{\rm G}(4)=\frac{\q}{\nu_c}=3\q= \frac{3d}{4}.
\label{Kim}
\end{equation}
Here, $\nu = 1/y_t^{\rm{G}} = 1/2$ is  the critical exponent governing the scaling of the correlation length with $t$ when $h=0$
and $\nu_c=1/y_h^{\rm{G}}(d_c)=2/(d_c+2)=1/3$ is that governing scaling in $h$ when $t=0$. 
Note that, unlike $y_t^{\rm G}$, the Gaussian value $y_h^{\rm G}$ and the renormalized values $y_t^*$ and $y_h^*$ are $d$-dependent.
Note also that the functions $F$ and $\Xi$ in Eqs.(\ref{eqfHomogeneity}) and (\ref{eqXiHomogeneity}) are clearly not the same as the respective functions in Eqs.(\ref{B11}) and (\ref{B22}), 
 but here, and throughout, we use the same notation loosely, to indicate a function of its arguments when precise knowledge of the function is neither available nor necessary.
Finally, it is also worth  noting that the dangerous variable does not modify the scaling dimension of the free energy density which is $d$ in Eq.(\ref{eqfHomogeneity}), as claimed explicitly in Ref.~\cite{BNPY}. 
That of the correlation length, on the other hand, becomes $\q$ in Eq.(\ref{eqXiHomogeneity}), as we argued before and showed explicitly in \cite{ourNPB},  correcting an explicit assumption made in Ref.\cite{BNPY} and elsewhere.

We  follow the ansatz made for the susceptibility in Ref.\cite{WiYo14} and  write the reduced temperature $t$ with respect to the pseudocritical  value $T_L$ of temperature $T$ in Eq.(\ref{eqfHomogeneity}) and elsewhere rather than to the critical point $T_\infty$, so that $t = T-T_L$.
For the $\phi^4$-theory, symmetry demands that the critical and pseudocritical values ($H_\infty$ and $H_L$, respectively) of the field $H$ must vanish so that $h = H-H_L \propto H$.
Hereafter, we omit arguments whose  reduced field vanishes. 
Explicitly, this means that scaling  in zero magnetic field, for example, is
\begin{eqnarray}
 f(L^{-1},t)&=&L^{-d}F\left[{L^{\frac{\qq}{\nu}}(T-T_L)}\right] ,
\label{eqfHomogeneity3} \\
 \xi(L^{-1},t)&=&L^{\qq}\Xi\left[{L^{\frac{\qq}{\nu}}(T-T_L)}\right] .
\label{eqXiHomogeneity3} 
\end{eqnarray}

As stated in the Introduction, $T_L$ may be defined by the temperature at which the correlation length of a finite system is maximum.
It differs from its infinite volume limit $T_\infty$ as 
\begin{equation}
t_\infty \equiv T_\infty-T_L \sim L^{-\lambda},
\label{eqlambda}
\end{equation} 
with 
$\lambda = 1/\nu$ for FBCs and $\lambda = \q/\nu$ for PBCs \cite{ourNPB}. 
For the correlation  length at pseudocriticality, 
this guarantees $\xi(L^{-1})\sim L^\qq =  L^{d/4}$, 
and for the free energy density $f(L^{-1})\sim L^{-d}$, 
irrespective of the boundary conditions. 
Equations (\ref{eqfHomogeneity3}) and (\ref{eqXiHomogeneity3}) 
are compatible with the specific-heat and correlation-length exponents 
$\alpha=0$ and $\nu=1/2$ for the $\phi^4$-model 
[from $f(t)\sim |T-T_\infty|^{\nu d_c} = |T-T_\infty|^{2}$ and $\xi(t) \sim |T-T_\infty|^{-\nu} =   |T-T_\infty|^{-1/2}$, respectively] 
if the scaling functions assume the asymptotic behaviour 
$F(x)\sim x^{\nu d_c}= x^{2}$ and $\Xi(x)\sim x^{-\nu}=x^{-1/2}$ for small $x$. 
At $T_\infty$, they accommodate  two distinct types of behaviour: 
$f(L^{-1},t_\infty) \sim L^{-4}$ and 
$\xi(L^{-1},t_\infty)\sim L$ for FBCs for which $\lambda=2$; and $f(L^{-1},t_\infty)\sim L^{-d}$ and 
$\xi (L^{-1},t_\infty)\sim L^{d/4}$ for PBCs for which $\lambda=d/2$.

This account (Subsection~\ref{History2}) completes our history of the standard picture of FSS above the upper critical dimension and of recent developments in $Q$-FSS theory. We have presented it in the context of $\phi^4$-theory. In Section~\ref{QFSSPerc} we bring $\phi^3$-percolation theory into the fold of $Q$-theory, resolving all of the above-mentioned outstanding issues above the upper critical dimension there. For convenience, in the next subsection we summarise some of the essential predictions of the new scheme.

\subsection{Summary of this paper}
\label{History3}

In Table~\ref{table1}, we summarise the essential outcomes of the unifying $Q$-FSS theory for percolation established in this paper in terms of the values of $\q$, $X$, $D$ and $\lambda$.
The critical exponents $\beta$, $\gamma$ and $\nu$, in the table, are  associated with the proportion of sites which are in the largest percolating cluster, the mean cluster size of the remaining clusters and the correlation length (also called connectedness length), respectively.
(Note that, below the upper critical dimension $\lambda$ is usually given by $1/\nu$ as in field theory, but this is not a requirement and one can construct boundary conditions which give alternative values of the shift exponent~\cite{JaKe02}. Other values of the shift exponent found experimentally can also result from irrelevant scaling variables below $d_c$ \cite{Andrieu98,Andrieu01}.
  Such alternatives arise through mechanisms different to that discussed in this paper, so we just give the usual value in Table~\ref{table1}.)

\begin{table}[t!]
\caption{Summary of finite-size scaling theory for percolation, expressed in terms of
$\xi \sim L^{\qq}$ for the correlation length, 
$N \sim L^X$ for the number of maximal clusters, the mass of which scale as $s_L \sim L^D$, 
and 
$|p_\infty - p_L|\sim L^{-\lambda}$ for the pseudocritical point. 
}
\vspace{0.25cm}
\begin{center}
{\begin{tabular}{lllll} 
 \hline \hline      
Regime of validity               
& $\q$                                   
& $X$
&  $D$ 
& $\lambda $    \\
 \hline   
 \vspace{0.1cm}
 $d<6$; at $p_L$ and $p_\infty$          
& $1$                                     
& ${\displaystyle{0 \quad}}$  
& ${\displaystyle{\frac{\beta+\gamma}{\nu}}}$                            
& ${\displaystyle{\frac{1}{\nu}}}$ (usually) \\
 \hline 
\vspace{0.1cm} 
$d>6$; FBCs  \begin{tabular}{l}at ${\displaystyle{ p_\infty}}$ \vphantom{${\displaystyle{\frac{\beta+\gamma}{\nu}}}$ }\\ ~ \\ at $p_L$ \vphantom{${\displaystyle{\frac{\beta+\gamma}{\nu}}}$ } \end{tabular}     
& \!\!\!\begin{tabular}{l} \\ ${\displaystyle{1\atop \vphantom 1}}$ 
                           \\  \\
                            ${\displaystyle{\frac{d}{6}\quad\quad}}$ \end{tabular}
& \!\!\!\begin{tabular}{l} ${\displaystyle{d-6 \quad}}$ \\ \\ \\
                     ${\displaystyle{0 \quad\quad}}$ \end{tabular} 
& \!\!\!\begin{tabular}{l} ${\displaystyle{\frac{\beta+\gamma}{\nu}=4}}$ \\ 
                     \\${\displaystyle{\frac{(\beta+\gamma)\q}{\nu}=\frac{2d}{3}\quad\quad}}$ \end{tabular}
&  ${\displaystyle{\frac{1}{\nu}=2}}$ 
\\  \hline
\vspace{0.2cm} 
$d>6$;   PBCs at $p_L,p_\infty$ & ${\displaystyle{\frac{d}{6}\quad\quad}}$ 
& ${\displaystyle{0 \quad}}$  
&${\displaystyle{\frac{(\beta+\gamma)\q}{\nu}=\frac{2d}{3}\quad\quad}}$
& ${\displaystyle{\frac{\q}{\nu}=\frac{d}{3}} }$    \\
\hline
	  \hline
\end{tabular}}
\end{center}
\label{table1}
\end{table}

Results in the fourth row for FBCs at $p_L$  
in Table~\ref{table1} are new, as are the scaling relations involving $\q$ in the  fourth and fifth columns as well as the unifying framework which gives consistency between disparate results discussed above.

In addition to the results laid out in  Table~\ref{table1}, 
which pertain to finite-size systems, 
one is interested in the density of size-$s$ clusters in infinite volume, which scales as $s^{-\tau}$ 
and the size of the largest clusters there, which scales as $|p-p_\infty|^{-1/\sigma}$. 
The associated scaling relations are $\tau = 2+\beta/(\beta+\gamma)$ and $\sigma = 1/(\beta + \gamma)$, 
and hold for all dimensions, below, at and above~$d_c$.
Above the upper critical dimension, where the mean-field critical exponents for percolation theory are 
\begin{equation}
 \beta  = 1, \quad \gamma = 1,  \quad \nu = \frac{1}{2}, \quad \nu_c=\frac{1}{4}, 
\label{MFTP1}
\end{equation}
these scaling relations give $\tau = 5/2$ and $\sigma = 1/2$.
The other mean-field critical exponents are, in standard notation, 
\begin{equation}
\alpha = -1, \quad  \delta = 2, \quad  \eta = 0.
\label{MFTP2}
\end{equation}

\section{$Q$-Scaling theory for percolation}
\label{QFSSPerc}

In this section we develop a heuristic $Q$-scaling theory for percolation.
Let us begin by recalling that percolation can be described by a $\phi^3$-expansion analogous to Eq.(\ref{eqLandauExpansion}) \cite{phi3perc1,phi3perc2}. 
The Gaussian fixed point is the usual one and the mean-field exponents of Eqs.(\ref{MFTP1}) and (\ref{MFTP2}) above can be deduced from the Gaussian fixed-point values:
\begin{eqnarray}
 \beta^{\rm G}(d)  & = & \frac{d-2}{4},    \quad
\gamma^{\rm G}(d)    =   1,                \quad 
   \nu^{\rm G}(d)    =   \frac{1}{2},         \quad
 \nu_c^{\rm G}(d)    =   \frac{2}{d+2},
\nonumber \\
\alpha^{\rm G}(d)  & = & 2-\frac{d}{2},       \quad
\delta^{\rm G}(d)    =   \frac{d+2}{d-2},  \quad
  \eta^{\rm G}(d)   =0,
\end{eqnarray}     
at $d=d_c=6$. 
Eqs.(\ref{eqfHomogeneity}) and (\ref{eqXiHomogeneity}) hold as before, with $T$, $T_L$ and $T_\infty$ replaced by $p$,  $p_L$ and  $p_\infty$, respectively. 
We henceforth denote the reduced probability $p-p_L$ by $t$. 
This vanishes at the pseudocritical point and its value at criticality is  $t_\infty = p_\infty - p_L$. 
The role of $H$ is played by the inverse of the cluster size $s$. 
Its pseudocritical value is not zero, however, and instead is given by the inverse of the largest cluster size which we denote by $s_L^{-1}$ (in analogy to $H_L$).
We then denote the reduced inverse cluster size  by $h = s^{-1}-s_L^{-1}$. 
Eq.(\ref{Eq-yGaussian}) also holds but Eq.(\ref{Kim}) has to be replaced by
\begin{equation}
y_t^*(d)=\q y_t^{\rm G}= \frac{\q}{\nu}=2\q=\frac{d}{3},
\quad{\rm and}
\quad 
y_h^*(d)=\q y_h^{\rm G}(6)= \frac{\q}{\nu_c}=4\q= \frac{2d}{3}.
\label{allofthem}
\end{equation}

The density per unit volume (or per site if the lattice spacing is taken to be unity) of clusters of size $s$ on a lattice of size $L$ is  $ n(L^{-1},t,h)$.
We follow Coniglio and Kadanoff-rescale the lattice lengths by a factor $b$~\cite{Conig85,Co00}.
We write
\begin{equation}
 L^\prime = \frac{L}{b}, \quad
 \xi^\prime = \frac{\xi}{b^\qq}, \quad
 s^\prime = \frac{s}{b^{D}}.
 \label{C5}
\end{equation}
(Refs.~\cite{Conig85,Co00} had an implicit assumption that $\q=1$ here.)
By comparing to the scaling forms of (\ref{eqfHomogeneity}) and (\ref{eqXiHomogeneity}), we identify
\begin{equation}
 D \equiv D(d) = y_h^*(d) = 4\q.
\label{Dstaryh}
\end{equation}
In percolation theory above the upper critical dimension $d_c=6$, therefore, $D = \q/\nu_c$ plays 
the same role as $y_h^*$ in the field theory.

Following Refs.\cite{Conig85,Co00}, if we assume that  clusters do not interpenetrate, 
their number  between $s$ and $s+ ds$ is the same before and after rescaling so that
\begin{equation}
 L^dn(L^{-1},t,h) ds = {L^\prime}^d n({L^\prime}^{-1},t^\prime,h^\prime) d s^\prime.
\label{interp}
\end{equation}
The combination $s n(L^{-1},t,h)$ then scales like
$  b^{-d}  F\left({{L^\prime}^{-1},t^\prime,h^\prime}\right)$,
and we identify this quantity as corresponding to the free energy density in Eq.(\ref{eqfHomogeneity}).

To study critical behaviour at the percolation transition, one defines a number of observables.
The first is
\begin{equation}
 K(L^{-1},t) = \sum_{s}{n(L^{-1},t,h)}.
\label{C01}
\end{equation}
As the percolation threshold is approached in the infinite-volume system,  
the singular part of $K(t)$ plays a role analogous to that of the singular free energy density of a spin  system:
\begin{equation}
 K(t) \sim  |p-p_\infty|^{2-\alpha}.
 \label{C1} 
\end{equation}
(As with the reduced fields, we drop $L^{-1}$ from the argument when it vanishes.)

We represent the probability that a site belongs to a percolating cluster  by $P(L^{-1},t)$.
The order parameter for the phase transition is then $P(t)$ 
and is equivalent of the spontaneous magnetisation in the case of a ferromagnetic-paramagnetic phase transition.
As such, it vanishes for $p<p_\infty$ and 
\begin{equation}
  P(t)  \sim   |p-p_\infty|^{\beta} \quad {\mbox{for}} \quad p>p_\infty.
 \label{C2} 
\end{equation}
Now, if a site is active, it is either in the infinite percolating cluster (of which there is only one for $p>p_\infty$) or in one of the finite  clusters, so~\cite{Stauffer,CM,FoAhCoSt04}
\begin{equation}
  p = P(t) + \sum_{s}{sn(t,h)},
\label{C02dash}
\end{equation}
where the sum excludes the single infinite percolating cluster.
A  finite-size counterpart to this equation is 
\begin{equation}
  p = P(L^{-1},t) + \sum_{s}{sn(L^{-1},t,h)},
\label{C02}
\end{equation}
where $P(L^{-1},t)$ is the probability that a site belongs to the maximal  clusters which are excluded from the summation.

The probability that an arbitrary site belongs to a finite size-$s$ cluster is $w(L^{-1},t,h) = s{n(L^{-1},t,h)} / \sum{s{n(L^{-1},t,h)}}$, 
where again the sum excludes the maximal  clusters.
Therefore the mean size of the remaining clusters selected by randomly choosing sites is
\begin{equation}
S(L^{-1},t) = \sum_{s}{sw(L^{-1},t,h)} = \frac{\sum_{s}{s^2 {n(L^{-1},t,h)}}}{\sum_{s}{s {n(L^{-1},t,h)}}} .
\label{C03}
\end{equation}
This corresponds to susceptibility for a magnet
and, in the infinite-volume limit,
\begin{equation}
 S(t)  \sim |p-p_\infty|^{-\gamma}.
 \label{C3} 
\end{equation}

While the quantities $K$, $P$ and $S$ correspond to the free energy and its moments, 
in percolation theory the infinite-volume correlation length $\xi (t)$ is the typical radius of the finite clusters. 
Again, the infinite cluster is excluded from the determination of $\xi (t)$ when $p>p_\infty$ \cite{CM}. 
The correlation length also diverges in the thermodynamic limit and  we write
\begin{equation}
 \xi(t)  \sim  |p-p_\infty|^{-\nu}.
 \label{C4}
\end{equation}
The finite-size counterpart is $\xi(L^{-1},t)$ and, from  Eq.(\ref{eqXiHomogeneity}), its rescaling is 
given by
\begin{equation}
\xi (L^{-1},t) 
 =
 b^{\qq}  \Xi\left({bL^{-1}, b^{{\frac{\qq}{\nu}}}{t}}\right).
\label{R4}
\end{equation}

Finally, we may use Eq.(\ref{C5}) to explicitly write the scaling for the number of sites in the largest cluster in a system of volume $L^d$ as~\cite{Conig85,Co00}
\begin{equation}
 s_L({t}) 
 = b^{D}\Sigma \left({bL^{-1}, b^{{\frac{\qq}{\nu}}}{t} }\right).
\label{Scaling-s}
\end{equation}
where $D = 4\q$, from Eq.(\ref{allofthem}).

As in the case of a thermal phase transition, there are scaling relations between the  five critical exponents $\alpha$, $\beta$, $\gamma$, $\nu$ and $\nu_c$.
Our aim is to understand the origin and status of these and other scaling relations in the percolation context as well as  the FSS of critical clusters.

If large clusters interpenetrate, Eq.(\ref{interp}) fails.
We account for this through an extra scaling factor so that~\cite{Conig85,Co00}
\begin{equation}
 L^dn(L^{-1},{t},{h}) ds = b^X {L^\prime}^dn({L^\prime}^{-1},{t}^\prime,{h}^\prime) ds^\prime,
 \label{Co6}
\end{equation}
where, as in Eq.(\ref{interp}), 
the primed variables are Kadanoff-rescaled according to Eq.(\ref{C5}).
 We re-express this as
\begin{equation}
 n(L^{-1},{t},{h})
 =
 b^{-(d-X+D)}  
 F \left({{bL^{-1}}, b^{{\frac{\qq}{\nu}}}{t}, {b^{\frac{\qq}{\nu_c}}} h }\right).
\label{C777}
\end{equation}
Comparing to  Eq.(\ref{eqfHomogeneity}), one sees that interpenetration of clusters is equivalent to a modification of the free energy density scaling dimension as a supplementary effect of the dangerous irrelevant variable, like the way it alters the correlation length scaling dimension. 
The scenario thus differs from that observed in $\phi^4$-theory where the free energy density scaling dimension was unchanged by the dangerous irrelevant variable. Hence, we expect that  hyperscaling violation in percolation manifests an additional feature compared to $\phi^4$-theory.

In developing a $Q$-scaling theory for percolation, we have to take into account the possibilities that 
(a) $\q \ne 1$  (b) $X\ne 0$.
Possibility (a) was dealt with in the field-theory context in Refs.~\cite{ourNPB,ourCMP,ourEPL,ourEPJB,ourreview,ourPRL} 
but (b) is a new feature to be incorporated into $Q$-theory here.
Similarly, possibility (b) was accounted for in Refs.~\cite{Conig85,Co00} but (a) is a new feature that needs to be incorporated into any account of scaling above the upper critical dimension.
We recognise that setting $\q=1$ recovers the starting point of previous theories above $d_c$~\cite{Conig85,Co00,Aizen97}.
In that case, Eq.(\ref{Dstaryh}) reverts to $D = y_h^{\rm{G}}(6) = 4$.

\subsection{Rescaling the correlation length}
\label{subsec1}

Setting $b=L$ in Eq.(\ref{R4}) we have
\begin{equation}
 \xi(L^{-1},t)=L^\qq\Xi \left({
                                            1, L^{\frac{\qq}{\nu}} t 
																				   }\right). 
\label{Eq21}
\end{equation}
The asymptotic form $\Xi(1,x)\sim x^{-\nu}$ for large $x$  guarantees  the recovery of Eq.(\ref{C4}).
At $p=p_L$ ($t=0$), Eq.(\ref{Eq21})  delivers ${\xi}(L)\sim  L^\qq$, with $\q=d/6$, consistent with Eq.(\ref{koppa}) 
and irrespective of boundary conditions.
 At $p=p_\infty$ ($t=t_{\infty}$), on the other hand, the asymptotic form 
\begin{equation}
 t_\infty = p_\infty - p_L \sim L^{-\lambda},
\end{equation}
from Eq.(\ref{eqlambda}),
 leads to
\begin{equation}
 \xi(L^{-1},t_\infty)\sim L^\qq\Xi \left({1,L^{\frac{\qq}{\nu}-\lambda}}\right) \sim L^{-\nu \lambda}.
\end{equation} 
As discussed in Subsection~\ref{History2}, according to $Q$-FSS, the value of $\lambda$  depends on the boundary conditions. 
In Ref.\cite{ourNPB} we found that  $\lambda = \q/\nu$ in the PBC case, so that 
${\xi}(L^{-1},t_\infty) \sim L^{\qq} $, i.e., the pseudocritical-type FSS extends as far as the critical point.
For FBCs, however,  $\lambda = 1/\nu$,  the shifting exceeds the rounding \cite{AhSt95,ourNPB,WiYo14}, and the asymptotic form $\Xi(1,x) \sim x^{-\nu}$ delivers $\xi(L^{-1},t_\infty) \sim L$. This is effectively given by Eq.(\ref{koppa}) with $\q$ set to $1$.  
Therefore, although  FSS at the pseudocritical point and FSS at the critical point come from the same scaling function in the FBC case (as they do in the PBC case), the $Q$-FSS window, which is  centred on the pseudocritical point, does not reach the critical point there  \cite{ourNPB,RuGa85,AhSt95,WiYo14,Italians}.

These results for PBCs recover a result in Refs.\cite{HeHo07,HeHo11}, where it was shown that, for sufficiently large $d$, and at the critical point $p_\infty$, percolation clusters on a torus of width $L$ are similar to those on a finite box with bulk boundary conditions, where the box has width $L^{d/6} \gg L$.
(Bulk boundary conditions are expected to lead to similar behaviour as FBCs at $p_\infty$.) 
Here we have shown that the same statement does not hold at the pseudocritical point; instead
FBCs and PBCs deliver the same (universal) $Q$-FSS for the correlation length at $p_L$.

\subsection{Rescaling maximal critical-cluster size}
\label{subsec2}
In Eq.(\ref{Scaling-s}), we again set $b=L$ to find
\begin{equation}
s_L(t)=L^{D}\Sigma \left({1,L^{\frac{\qq}{\nu}} t }\right).
\label{Scaling-s2}
\end{equation}
The scaling function admits the form $\Sigma(1,x)\sim x^{-1/\sigma }$, consistent with the thermodynamic limit
\begin{equation}
 s_\infty (p)\sim|p-p_\infty|^{-\frac 1\sigma}
\end{equation}
provided that
\begin{equation}
  \sigma = \frac{\q}{\nu D} = \frac{\nu_c}{\nu} = \frac{1}{2}.
\label{C9a}
\end{equation}
This comes from Eq.(\ref{Dstaryh}), irrespective of whether $\q=d/6$  or  $\q=1$.
For the finite-size value at the pseudocritical point, Eq.(\ref{Scaling-s2}) delivers
\begin{equation}
 s_L \sim L^{D}\Sigma(1,0)\sim L^{D}
\label{beerq}
\end{equation}
with $D=4\q = 2d/3$ irrespective of the boundary conditions.
At $p_\infty$, it yields 
\begin{equation}
 s_L(t_\infty)\sim L^{D}\Sigma \left({1,L^{\frac{\qq}{\nu}-\lambda}}\right)\sim L^{\frac\lambda\sigma}.
\label{crayons}
\end{equation}
If $\lambda = \q / \nu$ (which holds for PBCs), this is again $L^{D}$ with $D=4\q =2d/3$.
Therefore the fractal dimensionality of the largest cluster is $D=2d/3$ for PBCs at both the critical and the pseudocritical points, consistent with Refs.\cite{BoCh05a,BoCh05b,HeHo07,HeHo11}.
If $\lambda = 1 / \nu$ (which holds for FBCs), however, Eq.(\ref{crayons}) gives $D=4$  at $p_\infty$.
This result was suggested in Refs.\cite{AhGe84,Conig85,Aizen97,Co00} and proved rigorously in Refs.\cite{HeHo07,HeHo11} for bulk boundary conditions provided certain assumptions on $\eta$ hold \cite{HaSa90,HaHo03,FiHo15}.

Again, the new result derived here is that $s_L \sim L^D$ holds with $D=2d/3$ for FBCs at pseudocriticality.
We refer to this as random-graph-type asymptotics.

\subsection{Rescaling the cluster density}
\label{subsec2}

Let $b = h^{-\nu_c/\qq} = h^{-1/D}$ in Eq.(\ref{C777}) and write 
\begin{equation}
 h = s^{-1} - s_L^{-1} = s^{-1}\left({1-\frac{s}{s_L}}\right) 
\end{equation}
to obtain
\begin{equation}
 n(L^{-1},t,h) \sim
 s^{-\tau} {\mathcal{F}}\left({ s^{-\frac{1}{D}}L^{-1}, s^{\sigma} t }\right)
\label{red}
\end{equation}
to leading order in $s/s_L$ and where 
\begin{equation}
 \tau = \frac{d+D-X}{D}.
\label{C9}
\end{equation}
In the infinite-volume limit and at $p=p_\infty$, this gives that the density of size-$s$ clusters scales as
\begin{equation}
 n(h) \sim   s^{-\tau} .
\label{red2}
\end{equation}
Again,  $\tau$ has the universal value $5/2$.

\subsection{The number of  maximal clusters}
\label{subsec6}

From Eq.(\ref{C777}), we next set $b=L$ to  obtain the universal finite-size behaviour of the maximal cluster at $p_L$,
\begin{equation}
s_L n(L^{-1}) \sim L^{-(d-X)},
\label{43}
\end{equation}
while at $p_\infty$, one obtains
\begin{equation}
 s_L n(L^{-1},t_{\infty}) \sim L^{-(d-X)}F\left({1,L^{\frac{\qq}{\nu}-\lambda}}\right) \sim L^{-(d-X)},
\label{44}
\end{equation}
having used  $\lambda = \q / \nu$ with $\q=d/d_c$ in the PBC case and $\q=1$ for FBCs.

We identify the number of critical clusters scaling as $s_L$ as
\begin{equation}
N(L^{-1},t,h) = L^d 
      \int{ n(L^{-1},t,h) ds_L} 
			\sim 
			L^d s_L n(L^{-1},t,h) .
\label{cyan}
\end{equation}
This is the quantity which is analogous of the total free energy for a magnet.

From Eq.(\ref{43}) and (\ref{44}), we therefore identify 
\begin{equation}
N(L^{-1},t) \sim L^X
\label{magenta}
\end{equation}
at both {$t = t_\infty$} ($p = p_\infty$) and {$t = 0$} ($p=p_L$), so that $X$ is the exponent governing the number of critical and pseudocritical clusters. 
Following Refs.\cite{Conig85,Co00}, we assume that these are responsible for critical and pseudocritical behaviour so that
\begin{equation}
 K(L^{-1},t) \sim  L^{-d} N(L^{-1},t)  \sim L^{-(d-X)} \sim \xi(L^{-1},t)^{-\frac{d-X}{\qq}}
\label{blue}
\end{equation}
where {$t=t_\infty$ or $t=0$}
and we recall that  $\q$ is effectively $1$ in the former instance if FBCs are used.
Next, Eq.(\ref{C02}) gives
\begin{equation}
 P(L^{-1},t)  \sim \int{n(L^{-1},t,h) s_L ds_L} \sim  L^{-d} N(L^{-1},t) s_L \sim L^{-(d-X-D)} \sim \xi(L^{-1},t)^{-\frac{d-X-D}{\qq}},
\label{blue1}
\end{equation} 
which expresses the fact  that the probability for a  site to be  in any of the largest spanning clusters  equals the number of such clusters per unit volume times their size.  
Finally, Eq.(\ref{C03}), gives
\begin{equation}
S(L^{-1},t)  \sim \int{n(L^{-1},t,h) s_L^2 ds_L} \sim  L^{-d} N(L^{-1},t) s_L^2 \sim L^{-(d-X-2D)} \sim \xi(L^{-1},t)^{-\frac{d-X-2D}{\qq}}.
\label{blue2}
\end{equation}

Taking the thermodynamic limit, Eqs.(\ref{blue}), (\ref{blue1}) and (\ref{blue2}) become
\begin{eqnarray}
 K (t) \sim   &   \xi(t)^{-\frac{d-X}{\qq}} \quad &
                \sim    |p-p_\infty|^{\frac{\nu}{\qq}(d-X)}
 \label{C204a}\\
 P (t)  \sim  &    \xi(t)^{-\frac{d-X-D}{\qq}} &
                \sim    |p-p_\infty|^{\frac{\nu}{\qq}(d-X-D)}
 \label{C204b}\\ 
S (t)  \sim  &     \xi(t)^{-\frac{d-X-2D}{\qq}}&
               \sim     |p-p_\infty|^{\frac{\nu}{\qq}(d-X-2D)}.
 \label{C204c}
\end{eqnarray}
Comparing these with Eqs.(\ref{C1}), (\ref{C2}) and (\ref{C3}), we obtain
\begin{eqnarray}
 \alpha + 2 \beta + \gamma & = & 2,
\label{Co11}
\\
 D & = & \frac{\qq(\beta + \gamma)}{\nu},
\label{chock} \\
 \frac{\nu (d-X)}{\q} & = & 2 - \alpha.
\label{hhh}
\end{eqnarray}
The first of these is the Essam-Fisher relation~\cite{EsFi63}.
The second formula gives the fractal dimension of the largest clusters corrected for $Q$-theory.
The third is a hyperscaling relation. 
We will  show that it extends the traditional form of hyperscaling  to the {\emph{full}} scaling window  above the upper critical dimension (encompassing both the pseudocritical and critical points). 
Before that, we comment that Eqs.(\ref{C9a}) and (\ref{C9})  give
\begin{eqnarray}
\sigma & = & \frac{1}{\beta + \gamma}, 
\label{end2}
 \\
 \tau & = & 2 + \frac{\beta}{\beta + \gamma},
\label{end3}
\end{eqnarray}
each of which is independent of $\q$.
We now examine three scenarios for the $\q$-dependent Eqs.(\ref{chock}) and (\ref{hhh}). These are the most obvious scenarios we can think about, and they are realized in appropriate regimes discussed below, but we do not exclude other situations, e.g. crossover between them.\footnote{At the upper critical
 dimension itself ($d=d_c$), the dangerous irrelevant variables which give rise to $\q=d/d_c$ become marginal and logarithmic corrections ensue. 
In this case $\q$ becomes $1$, but a logarithmic counterpart arises, written as 
$\xi(L^{-1},t)\sim L (\ln{L})^{\hat{\qq}}$~\cite{ourEPL}. 
On the basis of the results collected in Ref.\cite{RKreview}, we conjecture that $\hat{\q} = 1/d_c$  is another scaling relation for logarithmic corrections, valid  in many models. 
For percolation theory at the upper critical dimension it delivers $\hat{\q} = 1/6$, for example. However, that the relation does not apply as an equality in all cases is clear from a small number of counterexamples such as in Ref.~\cite{Antonio}. 
Nonetheless, we may conjecture that there is a relation $\hat{\q} \propto 1/d_c$ where the constant of proportionality, although unknown in general, is frequently unity.}

\vspace{0.5cm}
\noindent
{\bf{Scenario 1:}} Eq.(\ref{hhh}) recovers the traditional form of hyperscaling ($\nu d = 2 - \alpha$) when  $X=0$ and $\q=1$.
This is the situation below the upper critical dimension and
Eqs.(\ref{chock}) then gives
\begin{equation}
 D =  \frac{\beta + \gamma}{\nu},
\label{end100}
\end{equation}
recovering the standard result there.

\vspace{0.5cm}
\noindent
{\bf{Scenario 2:}} When the traditional form of hyperscaling ($\nu d = 2 - \alpha$) fails and  $\q$ is effectively 1, we have the situation above the upper critical dimension considered in Refs.\cite{Conig85,Co00,Aizen97}. 
Eq.(\ref{end100}) holds there with $D$ ``stuck'' at $4$, independent of $d>6$,  as observed Ref.\cite{AhGe84}.
In this case, Eq.(\ref{hhh}) gives
\begin{equation}
 \nu (d-X) = 2 - \alpha,
\label{hhC}
\end{equation}
or
\begin{equation}
 X=d-d_c.
\label{d-X}
\end{equation}
Eq.(\ref{hhC}) means that $\nu X$ is an additive correction term, extending hyperscaling  above the upper critical dimension.
Eqs.(\ref{d-X})  means that the maximal clusters whose fractal dimension is $D=4$ proliferate  above $d=d_c$, just as  predicted in Ref.\cite{Conig85} and proved in Ref.\cite{Aizen97}.
But as we have seen,  the effective value $\q=1$   [and therefore Eq.(\ref{d-X})] only holds at the critical point $p=p_\infty$ for FBCs. 
This means that the scaling relation (\ref{hhC}) and the associated value of $X$ is non-universal.
We can refer to such proliferating clusters as {\emph{{spanning clusters}}} to follow previous terminology~\cite{Aizen97}.

\vspace{0.5cm}
\noindent
{\bf{Scenario 3:}} When $\q=d/d_c$ and
\begin{equation}
 X=0,
\label{x0}
\end{equation}
Eq.(\ref{hhh}) recovers Eq.(\ref{hyperscalingclusters}) in which $\q$ multiplicatively  extends hyperscaling above the upper critical dimension.
Eq.(\ref{x0}) means that the maximal critical clusters do not proliferate.
We already know this in the PBC case from Refs.\cite{BoCh05a,BoCh05b,HeHo07,HeHo11}.
What is new here is that this is valid universally whenever $\q = d/d_c$, which means it holds at the pseudocritical point for FBCs as well as at the pseudocritical (and critical) point for PBCs.
Eq.(\ref{chock}) gives
\begin{equation}
 D =  \q \frac{\beta + \gamma}{\nu} = \frac{2d}{3},
\label{end111}
\end{equation}
 meaning that $D$ is dimension dependent and is not ``stuck'' at its mean-field value $D=4$ as earlier thought~\cite{AhGe84}. 
To adhere to the terminology of previous theories, we may refer to these  non-proliferating clusters as  {\emph{{incipient} infinite clusters}}~\cite{IIF1,IIF2,IIF3}.

\vspace{0.5cm}

\subsection{Summary}

We can summarise the evolution of percolation theory over the past four decades in terms of the fractal dimensionality $D$ measured on the scale of the system extent $L$. 
The original renormalization-group-based scaling papers \cite{Conig85,Co00} gave $D=4$, a result that was made rigorous for bulk boundary conditions in Ref.\cite{Aizen97}. 
Further rigorous work \cite{BoCh05a,BoCh05b,HeHo07,HeHo11} established that $D$ takes a different value, namely $2d/3$ (random-graph asymptotics) for PBCs, both at $p_L$ and at $p_\infty$, severely reducing the scope of the original scaling theory~\cite{Conig85,Co00}.
Here we show that  random-graph-type asymptotics also hold for FBCs at $p_L$, perhaps a surprising result given that bulk  boundary conditions are expected to deliver the same behaviour as FBCs at $p_\infty$ so that $D=4$ there. 

We have seen that the {correlation length} is $\xi(L^{-1}) \sim L^\qq$ at pseudocriticality and also at criticality in the PBC case.
This is related to the {average} cluster mass $S_L \sim L^{\qq \beta / \nu} = L^{2\qq}$ via the usual expression of the susceptibility as the square of the correlation length above $d_c$~\cite{BoCh05a,BoCh05b,HeHo07,HeHo11}.
We have also seen that the {{largest}} critical clusters scale as $s \sim L^D = L^{2d/3}$.

The agreement between FSS at $p_L$ and at $p_\infty$ in the PBC case may be traced back to the fact that the shifting is of the same scale as the rounding there.  
This means that $p_\infty$, like $p_L$, is affected by dangerous irrelevant variables in the PBC case.
The difference between FSS at $p_L$ and at $p_\infty$ in the FBC case is because  the shifting exceeds the rounding there, so that $p_L$ and $p_\infty$ are in different scaling regimes.
Nonetheless, they are both governed by the Gaussian fixed point. Their difference comes from the fact that, while dangerous irrelevant variables are manifest at the pseudocritical point, $p_\infty$ is too far away to be influenced by them. FSS at $p_\infty$ remains Gaussian in its raw form  (effectively unadorned by dangerous irrelevant variables)~\cite{ourPRL}.
This is well established in $\phi^4$-theory [28-33,69] and implied from  early work on the subject \cite{RuGa85}. 
Here we have extended these considerations to $\phi^3$-theory. 

The difference between the shifting and rounding comes from the value of the exponent $\q$ which governs scaling of the correlation length.
Its role becomes clearer when we interpret critical clusters with $\q>1$ in terms of an unfolding procedure.
In particular, Eq.(6.23) of Ref.\cite{HeHo07}  shows that, after unfolding, typical points that are connected are of the order $L^{d/6}$ away from each other. 
In this sense, the fact that $\q=d/6$ is already present in the mathematics literature \cite{HeHo07,HeHo16},  for PBCs.
That $\q=d/d_c$ also appears for FBCs is the new claim of this paper.
One can interpret this as meaning  that the dimension of critical clusters  with $\q=d/6$ is again $4$ when  we measure distance on the correlation-length scale; 
on the scale of $\xi \sim L^{d/6}$, the volume  $L^D=L^{2d/3} = (L^{d/6})^4 = \xi^4$. 
In other words when measured on the scale of $\xi$, the fractal dimensionality of all critical clusters (both sets of boundary conditions, at both criticality and pseudocriticality) is universally $D=4$.
This brings us full circle in the field theory context and reconciles the various theories and predictions accounted for above.

\section{Discussion and Conclusions}
\label{Conclusions}

Here we presented a unified and consistent FSS picture for percolation theory by placing it within a recently developed renormalization-group-based framework for scaling above the upper critical 
dimension~\cite{ourNPB,ourCMP,ourEPL,ourEPJB,ourreview,ourPRL}.
The new picture stems from Eq.(\ref{R4}), which itself comes from extending the danger of irrelevant variables to the correlation sector in the renormalization-group approach. 
It does not require the notion of a thermodynamic length as discussed in various literature \cite{AhGe84,AnCoFo03,FoStCo04}. 
It also modifies the scaling picture of Eq.(\ref{C777}) for the free energy density.

The earlier explanation \cite{Conig85,Co00,Aizen97} for the breakdown of hyperscaling above $d_c$, namely the proliferation of  maximal clusters with fractal dimension 4 there, survives in $Q$-theory but only at the critical point for systems with FBCs, for which  $\q$ is effectively $1$. 
This picture does not apply at the pseudocritical point and therefore it is not universal.
Instead, the value of $\q$ is universally $d/d_c$ at the pseudocritical point 
in accordance with random-graph-type asymptotics.  
(Note that one has to distinguish between maximal clusters and spanning clusters - the scaling of the numbers of spanning clusters is expected to be the same for both sets of boundary conditions, both at criticality and pseudocriticality, so that they proliferate.)

This value for $\q$ explains the collapse of hyperscaling in its traditional form above $d_c$ and extends it via Eq.(\ref{hyperscalingclusters}) there.
In universal $Q$-theory, there is a multiplicative correction to hyperscaling [$\q^{-1}$ multiplies $d$ in Eq.(\ref{hhh})], in addition to the additive correction [$-X$ is added to $d$ in Eq.(\ref{hhh})] coming from non-universal older theories \cite{Conig85,Aizen97,Co00}.  
Although  the correlation length  scales faster than the system length, 
since the correlation volume is $\xi^{d_c}$, objects of length $\xi$ can fit into the lattice volume $L^d$ with $\q = d/d_c$~\cite{ourNPB}.
The folding of long clusters into cycles, which is a feature of percolation on Erd{\H{o}}s-R{\'{e}}nyi random graphs, is possible also for FBCs although, of course, considered as continuous curves, these are contractible, unlike those that wind around tori when PBCs are used \cite{HeHo16,HoSa14}. 

The $Q$-FSS prediction for the fractal dimensionality $D=2d/3$ of the largest clusters at pseudocriticality is universal and agrees with rigorous results established under certain conditions~\cite{Bo84,Lu90,Aldous,BoCh05a,BoCh05b}. 
The previous result $D=4$ for FBCs at the critical point~\cite{Aizen97} is based on an assumption, applicable to bulk boundary conditions at criticality, that the anomalous dimension $\eta$ vanishes above the upper critical dimension.
This holds on the scale of $L$ for FBCs at $p_\infty$~\cite{ourEPL,ourPRL}.
It is suggested in  Ref.\cite{ourEPL}  that this  also holds  on the scale of $\xi$ at the pseudocritical point.
Indeed, replacing $L$ by $\xi$ in the old theory of Refs.\cite{Conig85,Aizen97,Co00}, gives $s_L \sim \xi^4 \sim L^{2d/3}$, in agreement with Eq.(\ref{end111}).
This is the only length scale which survives in the infinite-volume limit, explaining bulk results there.

In this context too, we recall the discrepancy between the multiplicity of incipient spanning clusters and the uniqueness of the incipient infinite cluster. \cite{Aizen97}. 
These are considered as the same clusters  viewed at different scales~\cite{Aizen97}.
At very large (infinite) system size clusters so far apart that, from the point of view
of a site inside one of them, there is only one cluster  within a finite range.
But looking from outside, on a macroscopic scale, one can see many of them.  
We may identify the multiplicity of maximal clusters at the critical point in the FBC case as incipient spanning clusters.
The maximal clusters at the pseudocritical point, on the other hand, correspond to incipient infinite clusters in this framework.
This means that the identification of different FSS modes for the maximal clusters  in the FBC case that we suggest in this paper may offer a route to study incipient spanning and incipient infinite clusters separately, even experimentally (for spread-out percolation).

Percolation is a well-studied subject \cite{St97} with simple geometrical rules, robust continuous phase transitions and a vast variety of applications \cite{ArGr14}. 
So why were the new features identified here previously overlooked in the high-dimensional (or spread-out) versions?
In the study of critical phenomena below the upper critical dimension, choices between critical and pseudocritical points, as between boundary conditions, are normally immaterial for the leading FSS of systems of large enough size.
Although the sensitivity of the shift exponent to boundary conditions was appreciated in percolation theory above $d_c$ \cite{AhSt95,St94,StZi00}, the primacy of the pseudocritical point was not and computational practicalities meant many simulations focused on the critical point for free or mixed boundary conditions \cite{FoStCo04,St97,FoAhCoSt04,deArc87,RaKl88,AnCoFo03}.
Attempts to test statistical physicists' preferred theory, namely that based on renormalization-group ideas, which appeared to deliver proliferation ($X=d-d_c>0$) of the largest clusters (i.e., those with $D=4$) above $d_c$ \cite{Conig85,Aizen97,Co00} had limited success \cite{FoStCo04}.
Although this theory was supported by rigorous work, a focus on bulk boundary conditions rather than FBCs meant that mathematicians were also considering such systems at $p_\infty$ rather than $p_L$\cite{Aizen97}. 

In the meantime, and also for reasons of tractability, mathematicians focused on systems with PBCs 
\cite{FiHo15,HeHo16,BoCh05a,BoCh05b,HeHo07,HeHo11,HoSa14,NaPe08,KoNa11,IIF2,IIF3}. 
Although these are by now well understood, 
the fact that the largest clusters are characterised by  exponent values (namely $X=0$ and $D=2d/3$) different to those of the bulk-boundary-condition case seemed to indicate the absence of universality to FSS in percolation above the upper critical dimension. 
This was in line with the collapse of the standard form of hyperscaling the upper critical dimension (e.g., Ref.~\cite{PHA}).
Moreover, since the FSS form for PBCs was independent of the choice of $p_L$ or $p_\infty$, the expectation was reinforced that scaling at the FBC pseudocritical point would not deliver anything new. 

Here we have seen, however, that the pseudocritical point is key to linking the various sectors, for it is at $p_L$ and not at $p_\infty$ that universality resides \cite{ourNPB,ourCMP,ourEPL,ourEPJB,ourreview,ourPRL}. 
The existence of the non-trivial, universal, physical, pseudocritical exponent $\q$, which enters a corrected or extended hyperscaling relation, ensures that the largest clusters at $p_L$  are universally described by $X=0$ and $D=2d/3$ when measured on the scale $L$ of the underlying lattice. 
However, because of differences in rounding and shifting, the $Q$-scaling window is far more limited in the FBC case where, unlike for PBCs, it does not extend from the pseudocritical to the critical point. 
Instead, Coniglio's original theory holds at $p_\infty$ for FBCs~\cite{Conig85,Aizen97,Co00}.
Although universality does not hold on the scale of $L$ there, when measured on the correlation-length scale the fractal dimensionalities of all largest clusters is four, irrespective of which boundary conditions are used and of whether one sits at the critical or pseudocritical point. In  that sense, universality is restored in the entire scaling window.

As we have seen, throughout the long history of renormalization-founded approaches to the statistical physics of percolation theory, theoretical developments have been followed by attempts at numerical confirmation.
We hope that this paper will stimulate renewed efforts on the numerical side, clearly differentiating between criticality and pseudocriticality as well as between FBCs and PBCs. 
We hope, too, that it will stimulate rigorous studies to confirm or disprove the properties of high-dimensional percolation in the FBC case at pseudocriticality, which is the focus of our study.
 Finally, because of the connection between percolation and critical behaviour in the Ising model, for example, we suggest (see also Ref.\cite{AnCoFo03}) that the breakdown of hyperscaling in the latter case is also associated with rapidly diverging radii of Fortuin-Kasteleyn clusters, rather than only an infinite multiplicity of  maximal clusters.

\vspace{1cm}

\noindent
{\bf{Note Added in Proof:}}
Following submission of this paper, Juan Ruiz-Lorenzo published a paper revising an earlier estimate for the logarithmic correction to the correlation length for percolation in six dimensions \cite{RL1}. 
The new result agrees with our prediction that $\hat{\q} = 1/6$ and is contained in Ref.\cite{RL2}.

\vspace{0.5cm}
\noindent
{\bf{Acknowledgments:}} 
We thank Tim Ellis, Juan Ruiz-Lorenzo and Lo\"\i c Turban for  discussions and Ferenc Igl{\'{o}}i for comments at the start of this work.
We also thank Youjin Deng, Eren El{\c{c}}i, Emilio Flores-Sola, Tim Garoni, Jens Grimm and Martin Weigel for discussions.
 This work was supported by EU FP7 Projects 
No. 295302, ``Statistical Physics in Diverse Realisations,'' 
No. 612707, ``Dynamics of and in Complex Systems,'' 
and by the 
Doctoral College for the Statistical Physics of Complex Systems, 
Leipzig-Lorraine-Lviv-Coventry at the Franco-German University.

\bigskip
%

\end{document}